\documentclass[useAMS,usenatbib]{mn2e} % for a referee version
\usepackage{graphicx}
\usepackage{times}
\newcommand{\be}{\begin{equation}}
\newcommand{\ee}{\end{equation}}
\newcommand{\ba}{\begin{eqnarray}}
\newcommand{\ea}{\end{eqnarray}}

\newcommand{\mic}{\mu{\rm m}}
\newcommand{\mJy}{\rm mJy}
\newcommand{\uJy}{\mu{\rm Jy}}
\newcommand{\Msol}{\rm M_\odot}
\newcommand{\Lsol}{L_\odot}
\newcommand{\Msolpyr}{\rm M_\odot\,{\rm yr}^{-1}}
\newcommand{\kms}{\rm km\,s^{-1}}
\newcommand{\kpc}{\rm kpc}
\newcommand{\Mpc}{\rm Mpc}
\newcommand{\Myr}{\rm Myr}
\newcommand{\Gyr}{\rm Gyr}

\def\ls{\mathrel{\hbox{\rlap{\hbox{\lower4pt\hbox{$\sim$}}}\hbox{$<$}}}}
\def\gs{\mathrel{\hbox{\rlap{\hbox{\lower4pt\hbox{$\sim$}}}\hbox{$>$}}}}
\title[LIRGs in Abell 1758]{LoCuSS: Luminous infrared galaxies in the
  merging cluster Abell 1758 at $\bmath{z{=}0.28}$}

\author[Haines et al.]{C. P. Haines,$^{1}$ G. P. Smith,$^{1}$
 E. Egami,$^2$ N. Okabe,$^3$ M. Takada,$^4$ R. S. Ellis,$^{5,6}$ \and S. M. Moran,$^7$ K. Umetsu$^{8,9}$ \\
$^{1}$School of Physics and Astronomy, University of Birmingham, Edgbaston, Birmingham, B15 2TT, UK; cph@star.sr.bham.ac.uk\\
$^{2}$Steward Observatory, University of Arizona, 933 North Cherry Avenue, Tucson, AZ 85721, USA\\
$^{3}$Astronomical Institute, Tohoku University, Aramaki, Aoba-ku, Sendai, 980-8578, Japan\\
$^{4}$Institute of Physics and Mathematics of the Universe, The University of Tokyo, 5-1-5 Kashiwa-no-Ha, Kashiwa City, Chiba 277-8582, Japan\\ 
$^{5}$California Institute of Technology, 105-24 Astronomy, Pasadena, CA 91125, USA\\
$^{6}$Department of Astrophysics, University of Oxford, Keble Road, Oxford, OX1 3RH, UK\\
$^{7}$Department of Physics and Astronomy, The Johns Hopkins University, 3400 N. Charles Street, Baltimore, MD 21218, USA\\
$^{8}$Institute of Astronomy and Astrophysics, Academica Sinica, PO Box 23-141, Taipei 106, Taiwan\\
$^{9}$Leung center for Cosmology and Particle Astrophysics, National Taiwan
University, Taipei 10617, Taiwan
}

\begin{document}

\maketitle
\label{firstpage}
%\date{Accepted 1988 December 15. Received 1988 December 14; in original form 1988 October 11}

\begin{abstract}
  We present the first galaxy evolution results from the Local Cluster
  Substructure Survey (LoCuSS), a multi-wavelength survey of 100 X-ray
  selected galaxy clusters at $0.15{\le}z{\le}0.3$.  LoCuSS combines
  far-UV through far-IR observations of cluster galaxies with
  gravitational lensing analysis and X-ray data to investigate the
  interplay between the hierarchical assembly of clusters and the
  evolution of cluster galaxies.  Here we present new panoramic {\em
    Spitzer}/MIPS $24\mic$ observations of the merging cluster Abell
  1758 at $z{=}0.279$ spanning $6.5{\times}6.5\,\Mpc$ and reaching a
  $90\%$ completeness limit of $S_{24\mic}{=}400\uJy$.  We estimate a
  global cluster star-formation rate of ${\rm
    SFR}_{24\mic}{=}910{\pm}320\,\Msolpyr$ within $R{<}3\,\Mpc$ of the
  cluster centre, originating from 42 galaxies with
  $L_{8-1000\mic}{>}5{\times}10^{10}\Lsol$.  The obscured activity in
  A\,1758 is therefore comparable with that in Cl\,0024$+$1654, the
  most active cluster previously studied at $24\mic$.  The obscured
  galaxies faithfully trace the cluster potential as revealed by the
  weak-lensing mass map of the cluster, including numerous mass peaks
  at $R{\sim}$2--3\,\Mpc that are likely associated with infalling
  galaxy groups and filamentary structures.  However the core
  ($R{\ls}500\,\kpc$) of A\,1758N is ${\sim}2{\times}$ more active in
  the infrared than that of A\,1758S, likely reflecting differences in
  the recent dynamical history of the two clusters.  The 24$\mic$
  results from A\,1758 therefore suggest that dust-obscured cluster
  galaxies are common in merging clusters and suggests that obscured
  activity in clusters is triggered by both the details of
  cluster-cluster mergers and processes that operate at larger radii
  including those within in-falling groups.  Our ongoing far-UV
  through far-IR observations of a large sample of clusters should
  allow us to disentangle the different physical processes responsible
  for triggering obscured star formation in clusters.
\end{abstract}


\section{Introduction}
\label{intro}

\setcounter{footnote}{8}

Unravelling the physics of how gas-rich disk galaxies, recently
arrived in galaxy clusters, are transformed into the quiescent
elliptical and S0 populations that dominate dense cluster cores at
$z{\simeq}0$ is one of the longest-standing unsolved problems in
astrophysics. Numerous studies of massive early-type galaxies in
clusters have revealed predominately old stellar populations with
apparently very little growth in stellar mass in massive cluster
galaxies since $z{\sim}1$ \citep{tanaka,depropris,muzzin}.  This
indicates that the majority of stars in cluster early-type galaxies
formed at high redshift, $z{\gs}2$.  However \citet{bo78,bo84} found
that the fraction of cluster members with blue optical colours
increases from zero in the local universe to ${\sim}0.2$ by
$z{\simeq}0.4$, suggesting that the fraction of cluster galaxies that
are actively forming stars increases with increasing lookback time.
Empirically the star-forming spiral galaxies found by
\citet{bo78,bo84} at $z{\sim}0$.2--0.4 are mostly replaced by S0
galaxies in local clusters \citep{dressler80,dressler97,treu}.
Indeed, S0s may be completely absent at $z{\simeq}1$ \citep{smith},
although \citet{desai} suggest that the S0 fraction remains constant
over $0.5{<}z{<}1$ \citep[see also][]{postman}.  A simple
interpretation is that clusters accrete blue gas-rich star-forming
spirals at $z{\ge}0$.5--1 and that these galaxies were transformed
somehow into the passive S0s found in local clusters.

Numerous mechanisms have been proposed to deplete the reservoir of gas
in late-type spiral galaxies and thus cut off the fuel supply for
further star-formation, leading to transformation into S0 galaxies
\citep[for reviews see][]{boselli,haines07}.  Historically most of the
work on this area has concentrated on optical wavelengths, for example
identifying the so-called E+A galaxies \citep{poggianti}.  These
galaxies exhibit deep Balmer absorption lines and lack nebular
emission lines indicating that these apparently passive galaxies were
actively forming stars in the preceding ${\sim}0$.5--1.5\,Gyr. Indeed,
simple evolutionary models have been used to connect, via aging and
thus fading, this newly quiescent population with actively
star-forming spiral galaxies.

Mid-infrared observations of cluster galaxies with {\em ISO} and {\em
Spitzer} have challenged the idea that all spiral galaxies that fall
into clusters simply fade into passive galaxies via an intermediate
E+A phase.  For example, {\em ISO} revealed a population of luminous
infrared galaxies ($L_{IR}{\ge}10^{11}\Lsol$ -- LIRGs) and near-LIRGs
in clusters that imply that the star-formation rates derived from
optical diagnostics (e.g.\ [O{\sc ii}]) under-estimate the true star
formation rate (SFR) by ${\sim}1$0--3$0{\times}$ \citep[see][for a
review]{metcalfe}.  Deep radio observations have also revealed that
many E+As harbour significant ongoing star-formation missed by optical
diagnostics due to heavy dust obscuration
\citep{smail,miller03,miller06}.  The simple optically-derived fading
of spiral galaxies therefore seems to be substantially incomplete.

Radio and mid-infrared observations have also revealed an order of magnitude
cluster-to-cluster variations in the amount of obscured star formation
in clusters \citep{owen,fadda, duc02, duc04, metcalfe03, miller03, biviano, coia05a,
coia05b, geach, bai, marcillac, dressler08, saintonge, gallazzi}.
These differences have broadly been attributed to differences in the
recent dynamical history of the clusters \citep[e.g.][]{owen,miller03,geach},
however early {\em ISO} results had already suggested that a simple
one-to-one relationship between cluster-cluster mergers and obscured
star formation does not exist.  For example, {\em ISO} identified 10
LIRGs in Cl\,0024 and none in A\,1689, A\,2218, and A\,2390 (Coia et
al., 2005b; see also Metcalfe et al., 2005).  All four clusters are
well known strong lensing clusters, the mass distributions of which
are constrained by their strong lensing signal to be multi-modal,
i.e.\ indicative of recent cluster-cluster merging
\citep[e.g.][]{kneib,czoske,swinbank,limousin}.

Indications of the likely complex relationship between the dynamical
state of clusters and their current star-formation rate was also found
in the pioneering optical studies. For example, for the Coma cluster
\citet{caldwell} found numerous galaxies with post-starburst
characteristics (strong Balmer-line absorption) between the two X-ray
peaks corresponding to the main Coma cluster and the NGC\,4839
subcluster, suggestive of starbursts and subsequent quenching
resulting from a cluster-cluster merger. In contrast, \citet{tomita}
found no enhancement in the blue galaxy distribution between the two
X-ray peaks for an apparently similar system Abell 168. For this
latter case, \citet{tomita} suggests that the lack of enhanced
star-formation could be due to the galaxies being previously gas-poor
and hence unable to be triggered on interaction with their
environment.  \citet{bo84} measured the fraction of cluster galaxies
with blue optical colours out to $z{\simeq}0.5$, although their sample
was dominated by clusters at $0.15{\le} z{\le}0.3$.  The blue galaxy
fractions in these ``intermediate'' redshift clusters span the full
range of values found in clusters at $z{\simeq}0$ and $z{\simeq}0.5$,
i.e.\ $f_{\rm B}{\sim}$0--0.2.  Numerous UV/optical, infrared and radio
surveys have broadly confirmed that the scatter is large, however none
have combined the sample size and multi-wavelength dataset required to
relate directly actively star-forming galaxies in clusters to
differences in the intracluster medium and recent dynamical history of
clusters.

The Local Cluster Substructure Survey (LoCuSS; PI: G.\ P.\ Smith;
http://www.sr.bham.ac.uk/locuss) is a systematic multi-wavelength
survey of 100 galaxy clusters at $0.15{\le}z{\le}0.3$ drawn from the
ROSAT All Sky Survey cluster catalogues
\citep{ebeling98,ebeling00,bohringer}.  The overall goal of the survey
is to probe the relationship between the recent hierarchical infall
history of clusters \citep[as revealed by strong and weak lensing
observations;][]{smith08} and the baryonic properties of the clusters.
This results in two complementary aspects of the survey:
mass-observable scaling relations \citep{zhang08}, and evolution of
cluster galaxy populations.  Both of these themes are connected by the
aim of measuring and understanding the physical origin of the
cluster-to-cluster scatter.

This is the first in a series of papers about the first batch of 31
clusters observed by LoCuSS, for which the following data have been
gathered: wide-field Subaru/Suprime-CAM imaging, \emph{Hubble Space
Telescope} WFPC2 and/or ACS imaging of the cluster cores, wide-field
{\em Spitzer}/MIPS $24\mic$ maps to match the half degree Suprime-CAM
field of view, {\em GALEX} near- and far-ultraviolet (NUV/FUV)
imaging, and wide-field near-infrared (NIR) imaging obtained via a
combination of UKIRT/WFCAM and KPNO-4m/NEWFIRM.  We have also been
awarded 500ksec on {\em Herschel} as an Open Time Key Program to
observe this sample at $100$ and $160\mu m$ with PACS.  The galaxy
evolution goals that we will tackle with these data include to measure
robustly the scatter in the amount of obscured star formation in
clusters at ``low'' redshift \citep{haines09}, 
and to cross-correlate the location of
the obscured cluster galaxies with mass over-densities found in the
weak-lensing mass maps derived from the Subaru data \citep{okabe,okabe09}.
Full details of the survey design will be given in a future paper,
however here we emphasize that at the nominal redshift of our sample,
$z{=}0.2$, the Suprime-CAM field-of-view covers $6{\times}6\,\Mpc^{2}$
centred on each cluster.  A $10^{15}\Msol$ \citet{nfw} dark matter
halo has a virial radius of $r_{200}{\simeq}1.7\,\Mpc$.  Our survey
therefore probes out to $\sim1.5-2\times$ the virial radius of the
clusters.  The UV and IR data discussed above all cover at least the
same field of view as the Subaru data.  This survey is therefore
sensitive to the infall regions of clusters studied recently by, for
example, \citet{geach} and \citet{moran07}.

In this paper we present an analysis of Abell 1758 (A\,1758) at
$z{=}0.279$ as a case study on the influence of cluster-cluster
mergers on obscured star formation, and to outline the methods that
will be applied to the full sample in future papers.  A\,1758 was
selected for this study because previous X-ray and lensing studies
have identified it as a merging cluster, comprising two
gravitationally bound components -- one to the North and one to the
South (A\,1758N and A\,1758S respectively) -- both of which are
undergoing a merger \citep{david,okabe}. We note that throughout the
paper, we describe the star-formation measured from the $24\mic$ data
as obscured, making no distinction between heavily obscured
star-formation or the much less attenuated diffuse emission from
normally star-forming spirals as done by \citet{gallazzi}. The most
straightforward interpretation \citep{kennicutt07} of the $24\mic$
emission is that it traces the dust obscured star-formation, while the
observed UV or H$\alpha$ emission traces the unobscured one
\citep{calzetti}.  In future, we will combine UV and IR data to obtain
robust measures of both obscured and unobscured star-formation in
cluster galaxies, as well as the level of extinction.

In \S\ref{sec:data} we describe our observations of A\,1758, plus
archival data from the XMM-LSS field that we use to establish optimal
colour selection criteria to identify galaxies at the cluster
redshift.  In \S\ref{sec:analysis} and \S\ref{sec:results} we present
the photometric analysis and the main results respectively.  We
discuss the results in \S\ref{sec:discuss} and summarize them in
\S\ref{sec:conc}.  When necessary we assume \mbox{$\Omega_M=0.3$},
\mbox{$\Omega_\Lambda=0.7$} and \mbox{$H_0=70\kms\Mpc^{-1}$}, giving a
lookback time of $3.3\Gyr$ and an angular scale of $260\kpc$
arcmin$^{-1}$ at the cluster redshift.  All magnitudes are quoted in
the Vega system unless otherwise stated.

\section{Data}
\label{sec:data}

\subsection{Observations of A\,1758}\label{sec:a1758data}

The 24$\mu$m data were obtained with MIPS \citep{rieke} on
board the {\em Spitzer Space Telescope}\footnote{This work is based in part on
observations made with the Spitzer Space Telescope, which is operated
by the Jet Propulsion Laboratory, California Institute of Technology
under a contract with NASA (contract 1407).} \citep{werner}, and
consist of two separate AORs observed on January 6 and May 17, 2008
(PID:40872, PI G.\ P.\ Smith).  The pointing centres of the two AORs are
separated by 6$^{\prime}$, each covering a field of view of
25$^{\prime}{\times}25^{\prime}$ with a $5{\times}5$ grid.  In the earlier
AOR, the central pointing was excluded by using the fixed cluster
mode (with 400\arcsec\ offsets) since the area had already been imaged
by a Guaranteed Time Observations program 83 to a much deeper depth
($\sim$3000s/pixel).  However, the second AOR uses the raster mode
(i.e., with no gaps), and as a result, it contains the GTO area as
well.  For both the fixed-cluster and raster AORs, we performed at
each grid point two cycles of the small-field photometry observations
with a frame time of 3s, producing a per pixel exposure of 90s for
each AOR, or 180s in the overlapping region.

The 24$\mu$m data were reduced and combined with the Data Analysis
Tool (DAT) developed by the MIPS instrument team \citep{gordon}.
A few additional processing steps were also applied as described in
\citet{egami}.  The data were resampled and mosaicked with half
of the original instrument pixel scale (1\farcs245) to improve the
spatial resolution.

The clusters were observed with WFCAM \citep{wfcam} 
on the 3.8-m United Kingdom Infrared Telescope
(UKIRT)\footnote{UKIRT is operated by the Joint Astronomy Centre on
behalf of the Science and Technology Facilities Council of the United
Kingdom.} on March 10/20 2008 (ID: U/08A/32, PI: G.\ P.\ Smith).
A total exposure time of 800sec per pixel was accumulated in both
$J$- and $K$-bands using the same macrostepping ($13'$ offsets in RA
and Dec to cover the large gaps between the detectors), jittering and
$2{\times}2$ microstepping strategy used by the UKIDSS Deep
Extragalactic Survey \citep[DXS][]{ukidss}.  The data
were reduced and processed by the Cambridge Astronomical Survey Unit,
and photometrically calibrated to the Vega magnitude system using
2MASS stars in the fields.  The final pixel scale of the UKIRT data is
$0.2''/{\rm pixel}$; point sources detected in both filters have a full
width half maximum (FWHM) of $\sim1''$.

\begin{figure*}
\centerline{\includegraphics[width=180mm]{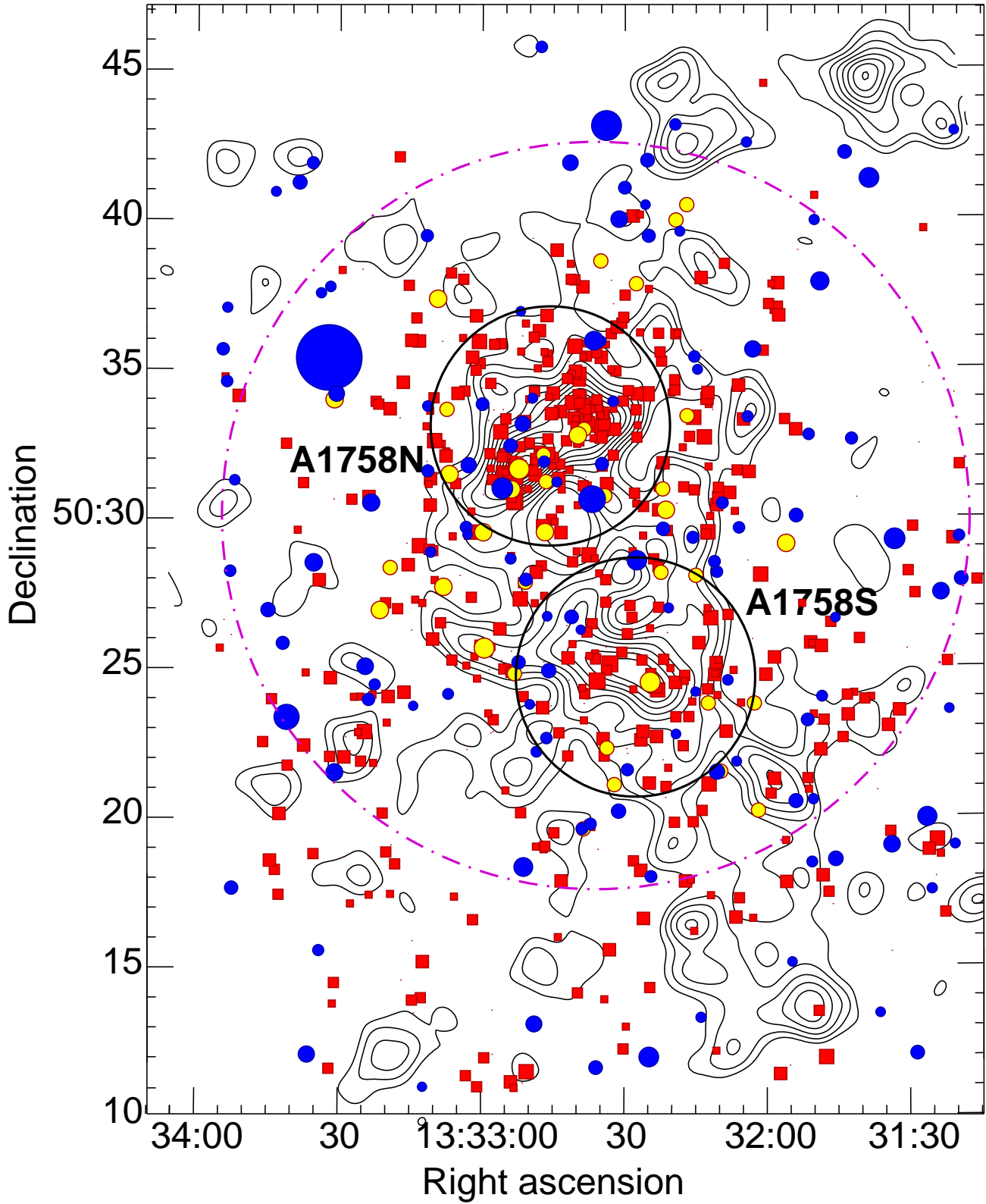}}
\caption{Spatial distribution of galaxies in the A\,1758 field whose
  $R'J'K'$ colours place them at the cluster redshift (see text). Red
  squares indicate passive spheroids ($n{>}2.5$ and no $24\mic$
  detection): the area of each symbol is proportional to the $K$-band
  flux. Yellow circles indicate passive spirals ($n{<}2.5$ and no
  $24\mic$ detection). Blue circles indicate star-forming disk
  galaxies; the area of each symbol is proportional to the $24\mic$
  flux. Overlaid are contours of the lensing $\kappa$-field
  reconstructed from weak shear data. The contours are spaced in units
  of $1\sigma$ reconstruction error. The two black circles indicate
  regions of 1\,Mpc radius centred on A1758N and A1758S, while the
  magenta dot-dashed circle indicates the 12.5\,arcmin ($3\Mpc$)
  radius region used to define the whole A\,1758 cluster population.
  }
\label{fig:spatial}
\end{figure*}
 
We also make use of wide-field optical imaging from the 8.3-m Subaru
Telescope\footnote{Based on data collected at the Subaru telescope,
which is operated by the National Astronomical Observatory of Japan.}.
These Suprime-CAM \citep{suprime} data (ID: S05A-159, PI: N.\ Okabe)
have been used by \citet{okabe} to study the mass distribution of
A\,1758 via the cluster's weak lensing signal -- the weak lensing mass
map from Okabe \& Umetsu is reproduced in Fig.~\ref{fig:spatial}.
Full details of the observations and data reduction are provided by
Okabe \& Umetsu.  The reduced data consist of stacked $g$ and $R_{C}$
(hereafter $R$) exposures totalling 720s in $g$ and 2880s in $R$.
Point sources in both filters have FWHMs of $\sim0.7''$.
Suprime-CAM's field of view is $34^{\prime}{\times}27^{\prime}$ with a
pixel scale of $0.206''/{\rm pixel}$.  Both optical frames were
registered onto the $K$-band frame using the {\sc iraf} tools {\sc
geomap} and {\sc geotran}. The transformations were defined by
comparison of the positions of 100--300 sources in each image, and
were found to be accurate to 0.1--0.$15''$ across the whole field.

\subsection{Archival XMM-LSS data}\label{sec:xmmlssdata}

In \S\ref{sec:colours} we use broad-band optical/near-IR colours of
$24\mic$ sources to identify them as likely cluster galaxies or not.
These selection criteria are derived from analysis of archival data on
the XMM-LSS field at $\alpha{=}2^{h}26^{m},\delta{=}{-}4.5^{\circ}$.
This field was observed at the relevant wavelengths by SWIRE
\citep[$24\mic$]{swire}, UKIDSS DXS \citep[$J,K$]{ukidss}, the CFHT
Legacy Survey D1 field ($ugriz$), and the VIMOS VLT Deep Survey
\citep[VVDS]{vvds}.

The properties of the SWIRE data are similar to our own $24\mic$
discussed in \S\ref{sec:a1758data}~\&~\ref{sec:phot}: an overall
exposure time per pixel of 80\,s, a final pixel scale of $1.2''$, and
complete to $\sim450\uJy$.  In our analysis we use the SWIRE
XMM-LSS DR2 catalogue from the entire 9\,deg$^{2}$ XMM-LSS
field\footnote{The SWIRE XMM-LSS DR2 catalogue is available here:
http://swire.ipac.caltech.edu/swire/astronomers/data\_access.html}.
The UKIDSS DXS $J/K$-band data also match our data well, with total
integration times of 640\,sec, the same macro- and micro-stepping
strategy, the same final pixel-scale, and comparable image quality of
${\rm FWHM}{\sim}0.9''$.  The final reduced UKIDSS DXS $J/K$-band
frames were obtained from the UKIRT science archive.  The CFHTLS D1
$g/r$-band frames were obtained from the CFHTLS Terapix First data
release at the Canadian Astronomy Data Centre.  These data have
slightly worse image quality than Okabe \& Umetsu's Subaru data (${\rm
FWHM}{\sim}0.9''$) and are $\sim0.5\,{\rm magnitudes}$ deeper.

Crucially for the calibration of our photometric selection methods
described in \S\ref{sec:colours}, a large spectroscopic redshift
catalogue is available for the XMM-LSS field.  The VIRMOS team obtained
8981 redshifts\footnote{available at
http://cencosw.oamp.fr/VVDS/VVDS\_DEEP.html} down to a limit of
$I_{AB}{=}24$, of which 1543 correspond to $K{<}19$ galaxies -- i.e.\
brighter than $\sim0.03L_K^\star$ at $z=0.28$.  In \S\ref{sec:results}
we consider only those with redshift quality flags of 2 or greater,
indicating a $>75\%$ confidence in the redshift.

\section{Photometric Analysis}
\label{sec:analysis}

\subsection{Object Detection and Multi-colour Photometry}
\label{sec:phot}

\begin{figure*}
\centerline{\includegraphics[width=170mm]{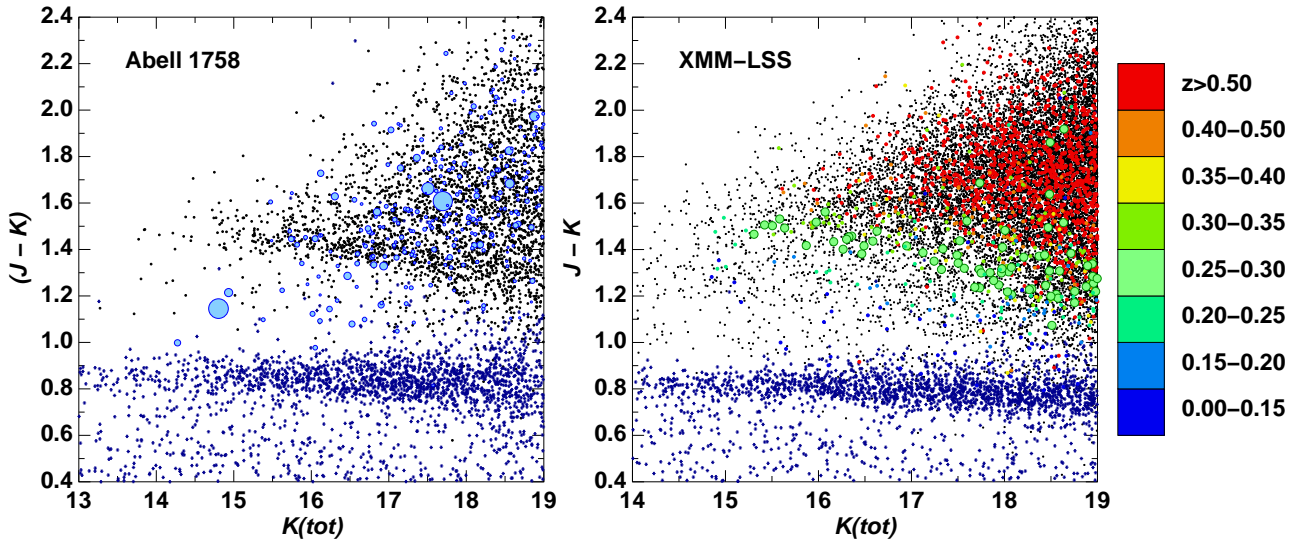}}
\caption{The $J-K/K$ C-M diagram of stars (blue stars) and galaxies
  (black points) in the Abell 1758 ({\em left panel}) and XMM-LSS
  ({\em right panel}) fields. In the {\em left panel} those galaxies
  detected by {\em Spitzer} are indicated by light-blue circles, whose
  area is proportional to their $24\mic$ flux. In the {\em right
    panel} galaxies with spectroscopic redshifts are indicated by
  symbols whose colour indicates their redshift from blue ($z{<}0.2$)
  to red ($z{>}0.5$) according to the key on the right. Those galaxies
  near the cluster redshift ($0.254{<}z{<}0.304$) are indicated by
  larger green symbols for emphasis. }
\label{vvds}
\end{figure*}

The $24\mic$ mosaic of A\,1758 was analysed with SExtractor
\citep{bertin}; following SWIRE we estimated the flux of objects
within an aperture of diameter $21''$, and applied an aperture
correction of $1.29$.  The flux detection limits and completeness of
the mosaic were determined by individually inserting 500 simulated
sources into the mosaic for a range of fluxes and determining their
detection rate and recovered fluxes, using identical extraction
procedures.  The sources used in the simulations were formed by
extracting isolated, high signal-to-noise and unresolved sources from
the mosaic itself. From these simulations, we estimate that the $90\%$
completeness limit of our $24\mic$ mosaic is $400\uJy$.

The optical and near-IR data on both A\,1758 and XMM-LSS were analysed
with SExtractor \citep{bertin} in two frame mode, such that objects
were selected in the $K$-band. The SExtractor {\sc mag\_auto}
magnitude was adopted for the $K$-band total magnitudes, and colours
were measured from the $g/R/J/K$-band frames within $2''$ diameter
apertures, after degrading the optical images to have the same FWHMs
as the $K$-band data.  Note that the XMM-LSS $r$-band photometry was
transformed to $R$ using the $(r-i)$ colour correction of Lupton
(2005)\footnote{http://www.sdss.org/dr6/algorithms/sdssUBVRITransform.html}.
The additional uncertainty introduced by this transformation is
0.0072\,mag.  These optical/near-IR catalogues were then used to
identify the optical/near-IR counterpart of each $24\mic$ source in
the A\,1758 and XMM-LSS catalogues, as the nearest $K$-band source
(where one exists) within $5''$ of the $24\mic$ centroid.  From the
completeness simulations described above we expect the positional
uncertainty of $24\mic$ sources to be $\sim0.75''$. In total 571
$S_{24\mic}{>}400\uJy$ sources were detected in the MIPS mosaic of
A\,1758.  Of these 455 have galaxy counterparts with $K{<}19$, 26 are
stars, and 90 have no $K{<}19$ counterpart within $5''$. These latter
sources we assume to be at high redshifts.

\subsection{Selection of Cluster Galaxies}\label{sec:colours}

\begin{figure*}
\centerline{\includegraphics[width=170mm]{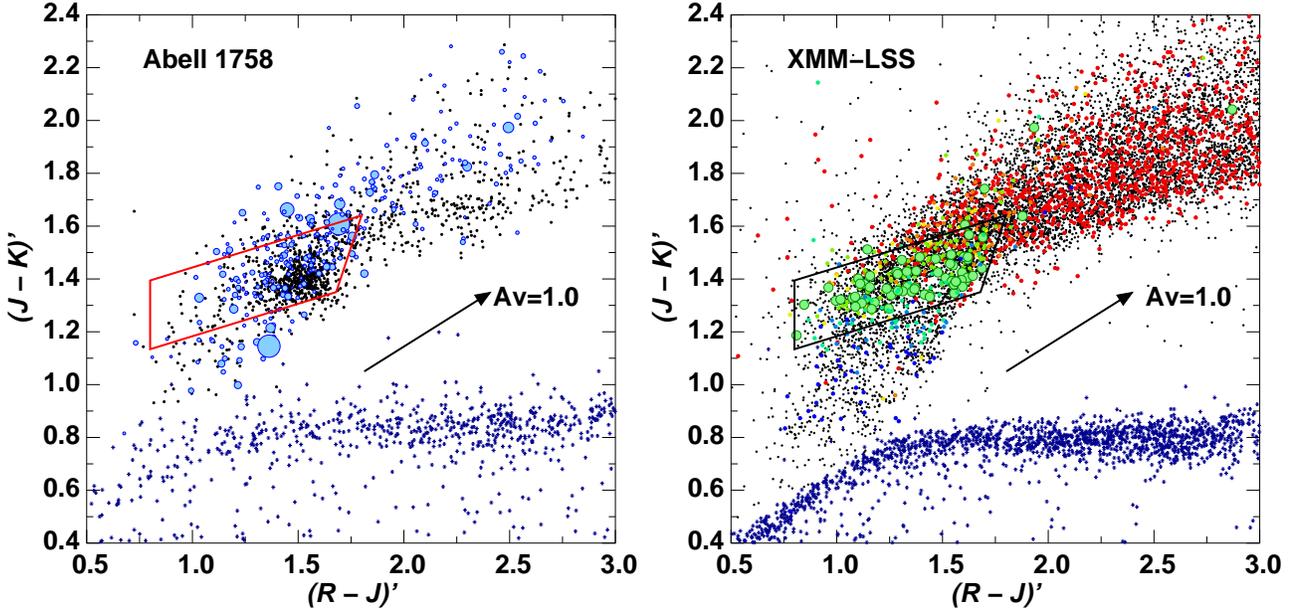}}
\caption{$R^{\prime}J^{\prime}K^{\prime}$ colour-colour diagram of stars and galaxies in
  the A\,1758 cluster field ({\em left panel}) and the XMM-LSS field
  ({\em right panel}). The $(R{-}J)^{\prime}$ and 
$(J{-}K)^{\prime}$
  colours are corrected to account for the slope of the C-M relation
  at $z{\sim}0.3$ (see text). Symbols in the {\em left} and {\em right
  panels} are as for the corresponding panels in Fig.~\ref{vvds}.  The
  black/red box indicates the colour selection criteria used to
  identify galaxies at the cluster redshift. The black arrow indicates
  the effect of dust extinction at the level of $A_{V}{=}1$\,mag on
  galaxy colours.  }
\label{a1758cols}
\end{figure*}

Galaxy colours depend on a complex combination of star-formation
history, metallicity, redshift and dust extinction, whose effects are
often difficult to disentangle, especially if only optical data are
available.  For example, it is particularly important to consider the
reddening effects of dust when considering IR-luminous sources whose
spectral energy distributions (SEDs) are likely to be strongly
affected by dust obscuration.  Fortunately, the SEDs of galaxies are
much less affected by both ongoing star-formation history and dust in
the near-IR than the optical; the $J-K$ colour varies by just
${\sim}0.1\,{\rm mag}$ across the entire Hubble sequence. In contrast
the $J-K$ colour increases monotonically with redshift to $z{\ga}0.5$,
from $J-K{\sim}0.9$ at $z{=}0$ to $J-K{\sim}1.6$ at $z{\sim}0.5$.
This near-IR redshift-colour relationship can be seen in
Fig.~\ref{vvds} -- galaxies of a particular redshift lie along a
single narrow colour-magnitude (C-M) relation, the slope of which is
due to the mass-metallicity relation whereby galaxies become
systematically more metal-rich with mass
\citep[e.g.][]{tremonti}. Note that we do not see a separate red sequence 
and blue cloud, due to the insensitivity of the $J-K$ colour to
current star-formation rate.

We exploit the fact that galaxies of a given redshift lie on such a
narrow C-M relation to empirically removing the effects of metallicity
on galaxy colour by fitting the slope of the C-M relation, and
subtracting it out.  We optimize this for the cluster redshift,
selecting those $K{<}19$ VVDS galaxies in the redshift range
0.254--0.304 (shown as large green symbols in the right panel of
Fig.~\ref{vvds}) and fitting their $J-K/K$ and $R-J/K$ C-M relations
using the biweight estimator, obtaining the following transforms:
\begin{eqnarray*}
(J-K)'&=&(J-K)+0.075\times(K-17.0)\\
(R-J)'&=&(R-J)+0.10\times(K-17.0).
\end{eqnarray*}
We show the resulting $R'J'K'$ colour-colour plot for A\,1758 and
XMM-LSS in Fig.~\ref{a1758cols}.  The robust separation of stars and
galaxies in $R'J'K'$ colour-space is striking; stars have much bluer
$J'-K'$ colours than galaxies. The only overlap between stellar and
galaxy loci is at $J'-K'{\sim}0.8$, $R'-J'{\sim}1.1$, i.e.\ for
galaxies at $z{\la}0.1$ which are of no interest in this study.
Star-galaxy separation is therefore achieved by selecting as galaxies
objects with {\em either} extended morphologies {\em or} $R'J'K'$
colours outside the stellar locus.  Using colour information to
perform star-galaxy separation instead of a conventional extended
versus unresolved morphological classification has the important
advantage of removing the potential bias against very compact
galaxies.

The large green symbols in the right panel of Fig.~\ref{a1758cols}
indicate those XMM-LSS galaxies at the cluster redshift
($0.254<z<0.304$), and can be seen to be well confined to a small
region in the $R'J'K'$ colour-space, and we use this to define a robust
colour-colour selection criterion of
\begin{eqnarray*}
(R-J)'&>&0.80\\
(J-K)'&>&0.94+0.24\times(R-J)'\\
(J-K)'&>&-2.71+2.42\times(R-J)'\\
(J-K)'&<&1.20+0.25\times(R-J)',
\end{eqnarray*}
as indicated in Fig.~\ref{a1758cols} by the black box.

We use the colours of those XMM-LSS galaxies with spectroscopic
redshifts to estimate the completeness and purity of the galaxy sample
selected using these colour cuts.  We find that the completeness of the
colour-selected sample is $90\pm4\%$ -- i.e.\ $90\%$ of XMM-LSS
galaxies at $0.254{<}z{<}0.304$ satisfy the colour cuts.  The sample
has a purity of $31\pm4\%$ -- i.e.\ $\sim30\%$ of the galaxies
selected in this way actually lie at $0.254{<}z{<}0.304$.  Varying the
colour cuts would adjust the trade-off between completeness and purity.
These colour cuts were chosen to optimise for completeness, with the
impurity being addressed via a statistical contamination subtraction
(\S\ref{sec:global}).  We find no evidence within the XMM-LSS data
that we are missing dusty star-forming galaxies at the cluster
redshift by not extending to redder $J'{-}K'$ or $R'{-}J'$
colours. Finally we note that objects with blue optical-NIR colours
($R'{-}J'<0.80$) are predominately quasars (not at the cluster
redshift), hence necessitating the blue cut in $R'{-}J'$.

We therefore apply these colour cuts to the A\,1758 photometric
catalogue, yielding 793 probable cluster members with $K{<}18$ and
within $3\,\Mpc$ of the cluster centre (left panel of
Fig.~\ref{a1758cols}) Note that the observed concentration of galaxies
with $R'-J'{\simeq}1.55$, $J'-K'{\simeq}1.4$, corresponding to the
early-type galaxies in A\,1758, provides independent confirmation of
the validity of the colour selection criteria developed above from the
XMM-LSS data.  The angular distribution of galaxies selected to be
likely cluster members in this manner is over-plotted on the
weak-lensing mass map in Fig.~\ref{fig:spatial}.  Hereafter we refer
to galaxies that satisfy the colour cuts developed in this section as
``cluster galaxies''.

 We searched NED\footnote{NASA/IPAC Extragalactic
     Database: http://nedwww.ipac.caltech.edu/} for galaxies with
   spectroscopic redshifts within the Subaru field of view.  Out of the
   25 galaxies with redshifts, 9 lie at $0.27{<}z{<}0.29$ and are
   therefore classified as cluster members.  These 9 include the BCG
   of both A1758N ($z{=}0.2792$) and
   A1758S ($z{=}0.2729$), but none of the galaxies that are detected
   at 24$\mu$m.  Nevertheless all 9 spectroscopically confirmed members
   satisfy our $R'J'K'$ colour selection criteria,
   confirming its high expected completeness.  Moreover, all eleven of
   the galaxies with $z{<}0.2$ were excluded from the cluster galaxy
catalogue on the basis of having too blue $J-K$ colours, confirming our
   ability to remove foreground galaxies.  We are less able
   to remove background galaxies, with 5 of 7 galaxies with
   $0.32{<}z{<}0.48$ classed as probable cluster members, an issue also
   apparent from analysis of the VVDS galaxies in the XMM-LSS field. In
   future articles, we will have much more complete redshift information
   both for this cluster and the other 30 clusters in the LoCuSS
   sample, through a recently commenced large spectroscopic programme
   to obtain redshifts for 200--400 members per cluster.

\section{Results}
\label{sec:results}

\subsection{Global Properties}
\label{sec:global}

We now turn to the population of dust-obscured galaxies in A\,1758: 82
of the 793 probable cluster members have $S_{24\mic}{>}400\uJy$ within
12.5\,arcmin ($3\Mpc$) of the centre of A\,1758, producing a combined
$24\mic$ flux of $85.8\mJy$.  We estimate the predicted contamination
of field $24\mic$ sources by using the optical/near-infrared
photometry of mid-infrared sources from the SWIRE XMM-LSS survey. In
total we identify 165 $K{<}18$ sources with $S_{24\mic}{>}400\uJy$
lying in the same $R'-J', J'-K'$ selection box used to identify
cluster members, within a total region of
2025\,arcmin$^{2}$. Correcting for this level of field contamination,
we report an excess of 42 mid-infrared sources with
$S_{24\mic}{>}400\uJy$, producing a combined flux of
$42.7{\pm}10.3\mJy$. The uncertainty in this last value, and the global measurements that follow, are estimated by assuming each 24$\mic$ source in the A1758 field has a probability of (42/82) of being a cluster member, and averaging over many Monte Carlo realizations of randomly selected cluster populations, such that the results are robust against any individual 24$\mic$ source turning out to be a cluster member or not. 

\begin{figure}
\centerline{\includegraphics[width=80mm]{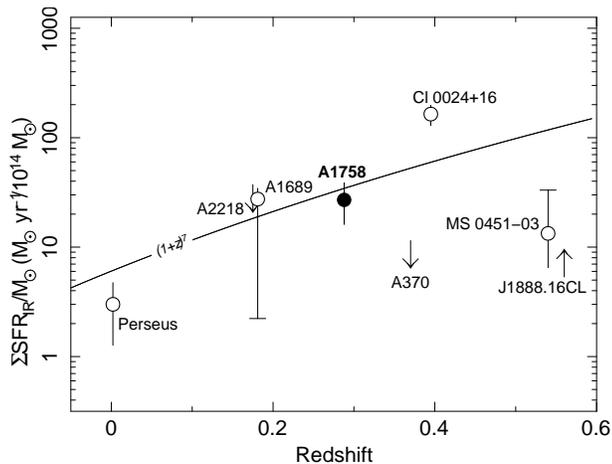}}
\caption{Variation in the specific SFR in clusters out to
  $z{\sim}0.5$. The SFRs are from the mid-infrared populations within
  ${\sim}3$\,Mpc, and these are normalized to the best estimate of the
  cluster mass, derived by weak lensing estimates or from the X-ray
  luminosities. The solid point corresponds to the estimate for
  A\,1758. The errors are derived from bootstrap resampling of the
  mid-infrared distribution. The open points and upper/lower limits
  for the remaining clusters are taken from \citet{geach}. Also
  plotted is an evolutionary model for the counts of star-forming
  ULIRGs from \citet{cowie}.}
\label{geach}
\end{figure} 
\setcounter{figure}{4}

Our $24\mic$ flux limit of $400\uJy$ corresponds to average total
infrared luminosities of $5{\times}10^{10}\Lsol$ in A\,1758.  This
translates into a star-formation rate of ${\sim}8.5\,\Msolpyr$,
assuming the star-formation calibration of \citet{lefloch}.  They
consider a range of SED templates to estimate the total infrared
luminosity, $L_{IR}{=}L($8--1$000\mic)$, of galaxies from their $24\mic$
flux, as a function of redshift. This calibration has a significant
amount of scatter due to the range of possible infrared SEDs, which at
$z{=}0.279$ corresponds to ${\sim}0.1-0.2$\,dex.  Integrating over
all the $24\mic$ sources within $3\Mpc$ of the cluster centre, we
estimate a global cluster star-formation rate of
$910{\pm}320\,\Msolpyr$, including a 0.1\,dex uncertainty in the
overall SFR calibration.  If we assign each individual cluster galaxy
to either A\,1758N or A\,1758S based on proximity to the respective
cluster centres then the estimated star formation rate of each cluster
is $580{\pm}210\Msolpyr$ and $330{\pm}60\Msolpyr$ respectively.
Following \citet{geach}, we also divide our global cluster SFR by the
cluster mass to estimate the specific star formation.  Adopting the
lensing results of \citet{okabe}, we estimate that the mass of
A\,1758, within the same $R<3\Mpc$ aperture used in this paper of
$3.2{\pm}1.5{\times}10^{15}\,\Msol$.  This mass translates to a
specific SFR of $29{\pm}16\,\Msolpyr/10^{14}\Msol$ placing A\,1758
among the more actively star-forming clusters (Fig.~\ref{geach}).

 In the above estimates we have neglected the possible contribution from AGN. 
We now investigate this by searching for X-ray point sources associated with the $24\mic$ sources in archive {\em XMM-Newton} and {\em Chandra} X-ray images, which cover both A1758N and A1758S \citep{david,okabe}, reaching sensitivities of $L_{X}($0.5--10\,keV$){\sim}10^{41.5}$\,erg\,s$^{-1}$ at the cluster redshift. Among our 82 probable cluster members with $S_{24\mu m}{>}400{\mu}$Jy we identify just two with coincident X-ray point sources, consistent with the typical X-ray selected AGN fractions in low-redshift ($z{<}0.3$) clusters of 1--5\% \citep{martini,silverman,gallazzi}. We note that this may be an underestimate of the AGN contamination, as many MIR-selected AGN are not detected in X-ray surveys \citep[e.g.][]{hickox}. In the future we will also be able to identify AGN from their optical spectra based on the emission-line ratios \citep{bpt}.

\begin{figure}
\centerline{\includegraphics[width=80mm]{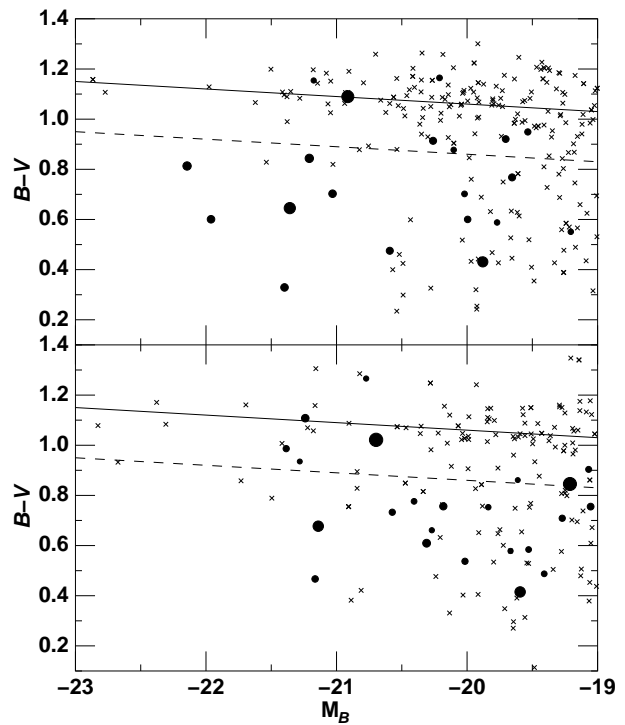}}
\caption{Rest-frame $B-V/M_B$ colour-magnitude diagrams for
  photometrically selected cluster members within 1\,Mpc of Abell
  1758N ({\em top panel}) and 1758S ({\em bottom panel}). Black circles
  indicate those with $S_{24\mic}{>}400\mu{\rm Jy}$, the area of each
  symbol being proportional to the $24\mic$ flux. Crosses indicate
  those galaxies not detected by {\em Spitzer}.}
\label{a1758cm}
\end{figure}
\setcounter{figure}{5}

\begin{figure}
\centerline{\includegraphics[width=80mm]{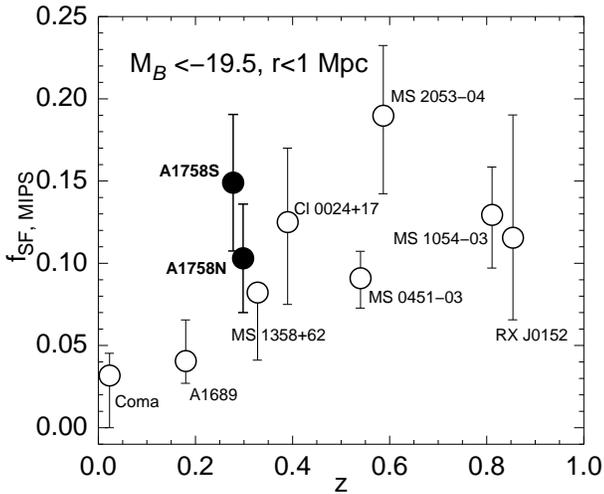}}
\caption{The mid-infrared Butcher-Oemler effect. The fraction of
  M$_{B}{<}{-}19.5$ cluster galaxies within 1\,Mpc of the cluster
  centre, that are star-forming as revealed by the MIPS $24\mic$
  observations, as a function of redshift. The solid symbols correspond
  to A\,1758N (lower) and A\,1758S (upper) at $z{=}0.279$ and are offset
  slightly in $z$ for clarity. The open symbols are taken from
  \citet{saintonge}.}
\label{bo}
\end{figure}

A final global measurement is the fraction of obscured star-forming
galaxies in a manner analogous to the pioneering measurements of blue
galaxy fractions by \citet{bo78,bo84}.  The advantage of making these
measurements in the mid-IR is that identifying galaxies as
star-forming solely on their optical colours is prone to
under-estimating the amount of star formation due to the obscuring effects
of dust in the most actively star-forming galaxies. Following
\citet{saintonge}, we estimate the fraction of cluster galaxies
brighter than $M_B{=}{-}19.5$ and located within 1\,Mpc of the cluster
centre that are detected by MIPS, $f_{SF,MIPS}$, separately for the clusters A1758N
and A1758S. For each cluster galaxy we derived rest-frame $B$-band
absolute magnitudes and $B-V$ colours from the observed Subaru $g$ and
$R$-band magnitudes. Following \citet{holden}, this was done by
deriving linear transformations from observed to rest-frame magnitudes
for model galaxies at the cluster redshift covering a range of
star-formation histories, using the stellar evolution code of
\citet{bc03}.  In Fig.~\ref{a1758cm} we show the resulting rest-frame
$B-V/M_B$ colour-magnitude diagrams for cluster galaxies within 1\,Mpc
of A\,1758N and A\,1758S.  Based on a comparison of source counts in
these regions and the XMM-LSS field, we expect that field
contamination to be at the level of one-in-ten for $M_B{<}{-}19.5$
galaxies, rising to one-in-four for those detected by {\em
Spitzer}. We see that most $24\mic$ sources would also be classified
as blue by the Butcher-Oemler criterion (i.e.\ blueward of the dashed
line in each panel), but a significant fraction (${\sim}20$\%) of the
star-formation in A\,1758 comes from galaxies with colours consistent
with those of early-type galaxies.  Taking into account field galaxy
contamination, we estimate the fraction of cluster galaxies brighter
than $M_B{=}{-}19.5$ and located within 1\,Mpc of the cluster centre that
are detected by {\em Spitzer} (i.e.\ $S_{24\mic}{>}400\uJy$) to be
$(10{\pm}3)\%$ for A\,1758N, and $(15{\pm}4)\%$ for A\,1758S
(Fig.~\ref{bo}).  By comparison to those galaxies in the XMM-LSS field satisfying the same colour selection criteria, we estimate that $(28{\pm3})\%$ of  $M_B{<}{-}19.5$ (assuming that they are at $z{=}0.28$) field galaxies at these redshifts have $S_{24\mic}{>}400\uJy$, somewhat higher than observed in the two clusters, as expected from the SF-density relation \citep[e.g.][]{lewis,haines07}.
 We note that in terms of $f_{SF,MIPS}$ A1758S appears the more active cluster, yet in terms of the global SFR, A1758N appears twice as active as A1758S, due primarily to A1758N being somewhat richer than A1758S.

\setcounter{figure}{6}

\begin{figure}
\centerline{\includegraphics[width=80mm]{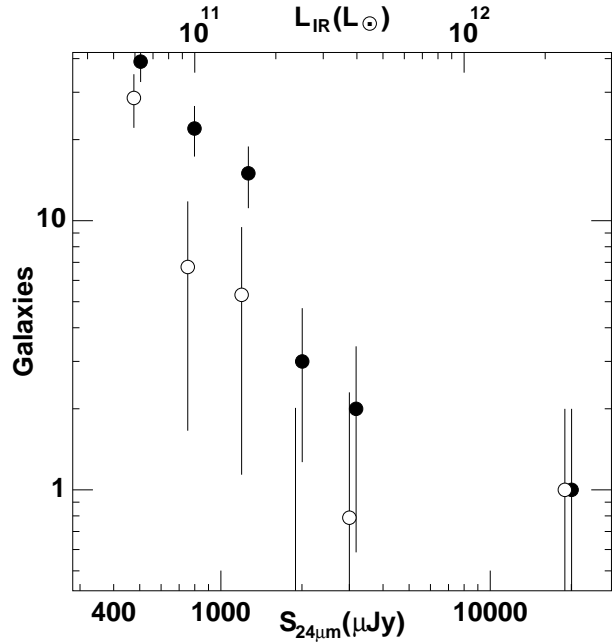}}
\caption{Number counts (filled symbols) and estimated luminosity
  function (open symbols) of $24\mic$ sources within 12.5\,arcmin of
  the A1758 cluster centre, and whose $R'J'K'$ colours place them at the
  cluster redshift. The $24\mic$ luminosity function is based on
  statistical subtraction of $24\mic$ sources in the XMM-LSS field
  satisfying the same colour criterion.  Scaled estimates of the
  bolometric infrared luminosities, based on the calibration of
  \citet{lefloch} are indicated along the top axis.  }
\label{fig:lf}
\end{figure} 

\subsection{Number Counts}
\label{sec:counts}

In Figure~\ref{fig:lf} we show the differential number counts (solid
black points) of $24\mic$ sources with $R'J'K'$ colours consistent
with the cluster redshift and within $3\Mpc$ of the cluster centre. The
open points show the differential 24$\mic$ number counts after
correcting for field galaxy contamination on the basis of the $24\mic$
sources in the XMM-LSS field satisfying the same colour selection,
providing an estimate of the $24\mic$ luminosity function for
A1758. According to this, almost all of the cluster $24\mic$ sources
have $S_{24\mic}{<}1600\uJy$, which corresponds to a bolometric
infrared luminosity at $z=0.279$ $L_{IR}{<}2{\times}10^{11}\Lsol$
based on the calibration of \citet{lefloch}. We estimate there to be
$10\pm6$ LIRGs in Abell\,1758, the same number as found for Cl\,0024,
although clearly without any redshift information this number is quite
uncertain. We also identify one possible cluster ULIRG
($L_{IR}{>}10^{12}\Lsol$), which appears from the optical image to be a
merging galaxy system with a combined NIR luminosity of
${\sim}0.4\,L_{K}^\star$. 

\subsection{Surface Photometry and Morphology}
\label{sec:morph}

We now turn to the morphology of the dust-obscured galaxies in
A\,1758.  Ideally this analysis would be based on wide-field high
resolution imaging with the \emph{Hubble Space Telescope}, allowing
detailed morphological classifications across the full field
\citep[e.g.][]{moran07}.  In the absence of such data, we analyse the
Subaru $R$-band frame, using {\sc 2dphot} \citep{labarbera} to derive
structural parameters for each cluster galaxy.  {\sc 2dphot} is an
automated tool to obtain both integrated and surface photometry of
galaxies, the main steps of which are: i) estimation of the FWHM and
the robust identification of stars; ii) construction of an accurate
PSF model, taking into account both possible spatial variations as
well as non-circularity; iii) derivation of structural parameters
(effective radius $r_e$, mean surface brightness ${<}\mu{>}_e$, and
Sersic index $n$) by fitting 2D PSF-convolved Sersic models to the
data.  The high-quality Subaru data, obtained in excellent conditions,
allows us to obtain structural parameters accurate to
${\delta}n/n\sim0.05$.

\begin{figure}
\centerline{\includegraphics[width=80mm]{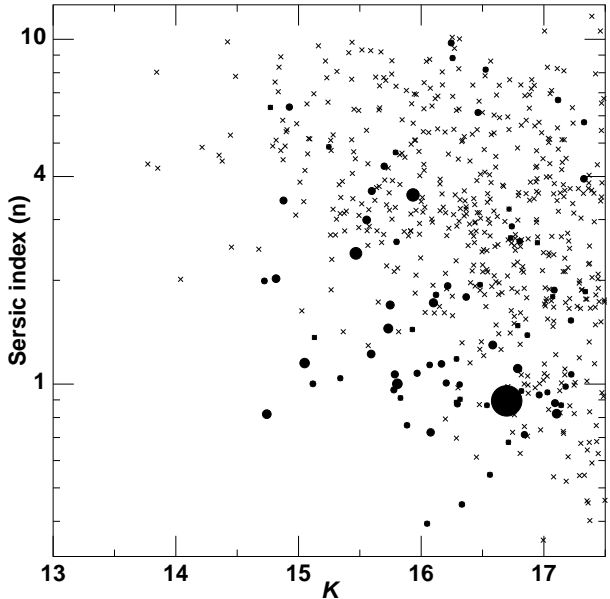}}
\caption{Morphologies and $K$-band magnitudes of those galaxies within
  12.5\,arcmin of the A1758 cluster centre whose $RJK$ colours place
  them at the cluster redshift. Circles indicate galaxies
  detected with {\em Spitzer}, the area of each symbol is proportional
  to the $24\mic$ flux. Crosses indicate those galaxies not
  detected in $24\mic$.}
\label{fig:morph}
\end{figure} 

In Fig.~\ref{fig:morph} we plot Sersic index $n$ versus $K$-band
magnitude for all cluster members.  Those galaxies detected by {\em
Spitzer} are indicated by circles whose area is proportional to the
$24\mic$ flux, while non-detections ($S_{24\mic}{<}400\mic$) are shown
as crosses. Sersic indices can be used to achieve a crude
morphological classification into bulge- and disk-dominated galaxies
\citep[e.g.][]{driver,ball}, and so adopting a nominal dividing line
of $n{=}2.5$ we classify the {\em Spitzer}-detected cluster galaxies
as being either bulge-dominated ($n{>}2.5$) or disk-dominated
($n{<}2.5$).  This yields the following fractions:
$(26.8^{+11.1}_{-9.4})\%$ of mid-IR-bright cluster galaxies are
bulge-dominated and $(73.2^{+9.4}_{-11.1})\%$ are disk-dominated where
the quoted uncertainties are $2\sigma$ binomial error bars following
\citet{gehrels}.  The {\em Spitzer}-detected cluster galaxies are
therefore dominated by disk galaxies, indeed many have large axis
ratios which may suggest that some of the dust obscuration is caused
by viewing these galaxies close to edge-on. However, we see no
significant difference in the axis ratios between those disk galaxies
detected by Spitzer and those not, which points against any major
inclination effect. Although a quarter of mid-infrared bright cluster galaxies are early-types, these galaxies contribute just ($13\pm4\%$) of the total 24$\mic$ flux estimated in \S\ref{sec:global}. Hence, even if all of the 24$\mic$ emission in these galaxies is due to AGN, they do not significantly effect our global SFR estimates in \S\ref{sec:global}.

\subsection{Spatial Distribution}
\label{sec:spatial}

In Figure~\ref{fig:spatial} we over-plot the spatial distribution of
cluster galaxies in A\,1758 on the projected cluster mass distribution
from Okabe \& Umetsu (2008; black contours).  The spatial distribution
of cluster early-type galaxies (filled red squares) closely follow the
underlying dark matter distribution.  The cores of both of the merging
components of A\,1758N (at $\alpha{\simeq}13{:}32{:}50$,
$\delta{\simeq}50{:}32$ and $\alpha{\simeq}13{:}32{:}40$,
$\delta{\simeq}50{:}33.5$ -- see \S\ref{sec:discuss}) are
particularly dominated by early-type galaxies.  Numerous mass peaks of
lower significance e.g.\ at $\alpha{\simeq}13{:}33{:}30$,
$\delta{\simeq}50{:}23$; $\alpha{\simeq}13{:}31{:}40$,
$\delta{\simeq}50{:}23$; and $\alpha{\simeq}13{:}32{:}00$,
$\delta{\simeq}50{:}21$), also contain over-densities of early-type
galaxies.

The distribution of dusty star-forming galaxies ($S_{24\mic}{>}400\mic$;
blue circles) is much less concentrated spatially than the
early-types, consistent with being an infalling population.  An
important caveat is that ${\sim}50\%$ of these galaxies are probably
field contaminants that should be randomly distributed across the
field.  Indeed, these field galaxies may help to explain the
population of star-forming galaxies that are not associated with
mass-over-densities in the weak-lensing mass map.  On the other hand
most of the dusty galaxies are indeed associated with over-densities
in the mass map.  In addition to the population of dusty disk galaxies
that are associated with mass over-densities at large radii --
presumably in-falling galaxy groups -- we also find dusty disk
galaxies occupying the central region (projected radii of
$R{\ls}500\kpc$) of A\,1758N.  In contrast the central region of
A\,1758S is completely devoid of dusty disk galaxies.

\begin{figure}
\centerline{\includegraphics[width=80mm]{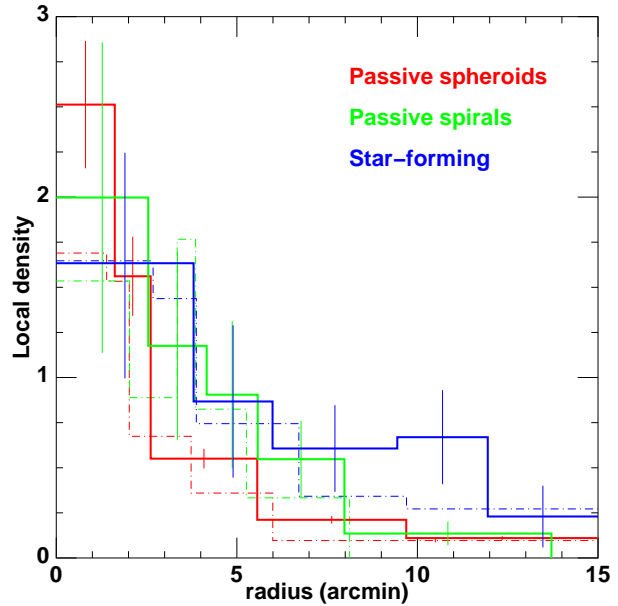}}
\caption{Comparison of the radial distributions of passive spheroids
  (red histograms), passive spirals (green histograms) and
  star-forming galaxies (blue histograms) in the A\,1758 cluster
  field. The solid histograms indicate clustercentric radii from just
  A1758N, while the dot-dashed histograms take the radius to be
  distance of each galaxy from the nearest of A1758N and A1758S. In
  each case, we correct the densities for field contamination based on
  the XMM-LSS field. The y-axis is an arbitrary density scale,
  normalized for ease of comparison between the galaxy populations.  }
\label{fig:radial}
\end{figure} 

We also identify a third significant population of galaxies in A\,1758
-- passive spiral galaxies, defined empirically by: $K{<}16.5$,
$n{<}2.5$ and undetected at $24\mic$, i.e.\ $S_{24\mic}<400\uJy$
(yellow circles in Fig.~\ref{fig:spatial}).  We identify 37 such
galaxies all of which lie within 3\,Mpc of the centre of either
A\,1758N or A\,1758S.  This population trace the underlying mass
distribution very well, and appears to be more concentrated than the
dusty spiral galaxies, but less concentrated than the early-type
galaxy population.  The apparent concentration of these objects
towards the cluster centre, suggests that these are a cluster
population.  This is confirmed by their relative rarity in field
regions -- we find a density of just $34{\pm}13\,{\rm deg}^2$ such
galaxies in the XMM-LSS data, corresponding to an expected field
contamination of just $4.3{\pm}1.6$ within a region of projected
radius $R{=}3\,\Mpc$.

To examine the relative concentrations of the three galaxy populations
discussed above more quantitatively, we show in Fig.~\ref{fig:radial}
the radial distributions of early-type galaxies, passive spirals, and
dusty star-forming galaxies in A\,1758, after correcting for field
galaxy contamination.  The number density of all three populations
decreases with increasing projected cluster-centric radius, confirming
that all three are associated with the cluster.  The distribution of
late-type galaxies (whether detected or not at $24\mic$) is less
centrally concentrated than the distribution of early-type galaxies;
the radial distributions of $24\mic$-detected and undetected late-type
galaxies are consistent with the errors out to $\sim8'$, however at
larger radii there is an excess of $24\mic$-detected galaxies over
both early-types and passive spirals.  This excess corresponds to the
galaxies discussed above that are associated with galaxy groups
detected in the out-skirts of the clusters in the weak-lensing map.

\section{Discussion}\label{sec:discuss}

\subsection{A\,1758 as a Cluster Merger -- The Mid-IR View}

\citet{okabe} and \citet{david} have studied A\,1758 in detail as a
merging cluster using weak lensing and X-ray data respectively.  In
particular \citeauthor{david} investigated the thermodynamics of both
cluster components.  Here we compare our results on obscured
star-formation in A\,1758 with the thermodynamics of the intracluster
medium, and thus with the likely dynamical state of the cluster.

\citeauthor{david} found no evidence of compressional heating of the
intracluster medium (ICM) in the region between A\,1758N and A\,1758S,
suggesting that the two clusters have not begun to merge with each
other yet, although they may be gravitational bound to each other
already.  A\,1758N itself is interpreted as being a large impact
parameter merger between two 7\,keV clusters, one cluster (to the
North-West) currently moving to the North and the other (to the
South-East) currently moving to the South-East.  In contrast, the
merger in A\,1758S is interpreted as having a much smaller impact
parameter with the merger occurring close to the line-of-sight through
the cluster.  The core radii of the merging components within A\,1758N
are $\sim100\kpc$; in contrast the merging components of A\,1758S have
core radii of $200\kpc$ and $350\kpc$. \citeauthor{david} therefore
suggested that, as a consequence of the differing impact parameters, the galaxies in the merging clusters of A\,1758S have experienced lower ram pressures than those in A\,1758N, and that A\,1758S may be at an earlier
stage in its merger than A\,1758N.  However, both A\,1758N and
A\,1758S do contain regions of hotter gas that have presumably been
shock-heated during the respective mergers.  The precise timing
difference between the two mergers is difficult to ascertain, however
\citet{david} proposed that A\,1758N is seen significantly after the
point of closest approach whilst A\,1758S may be seen at approximately
the epoch of closest approach.  Okabe \& Umetsu's weak lensing mass
map successfully resolves the two merging components of A\,1758N, and
locates a single mass peak in the core of A\,1758S, both of which are
consistent with David \& Kempner's interpretation of the X-ray data.

How well do our mid-IR results fit with the X-ray and lensing results?
One of the most striking features of the distribution of obscured
galaxies in Fig.~\ref{fig:spatial} is their absence from the central
$R\ls500\kpc$ region of A\,1758S, in contrast to the same region in
A\,1758N, especially in the vicinity of the South-East component of
A\,1758N.  This difference between the clusters is reminiscent of the
large cluster-to-cluster scatter within the core regions of merging
clusters found with {\em ISO} \citep{metcalfe}.  Intense star
formation can be triggered in cluster-cluster mergers by the passage
of gas-rich galaxies through shocked regions of the ICM
\citep{roettiger}.  In A\,1758N the merger-induced shock-front(s) have
dissipated \citep{david}, which helps to explain why shock-fronts are
not found adjacent to the dusty galaxies in the core of that cluster.
A\,1758S is claimed to be an earlier stage in the merging process,
however \citeauthor{david} did detect shock-heated gas in this
cluster.  Therefore naively one would expect that star-formation would
have been triggered by the shocks as in A\,1758N.  However no $24\mic$
sources are found in the central $R{\ls}500\kpc$ region of A\,1758S.
The merger geometry of A\,1758S may help to explain the absence of
obscured galaxies in its core.  For example the $24\mic$ sources at
$R\sim500-1000\kpc$ may have been scattered to large radii by the
line-of-sight merger that is ongoing in the core of the cluster in a
manner analogous to the distribution of cluster galaxies in Cl\,0024,
another line-of-sight merger cluster \citep{czoske}.  Assuming that
the galaxies are moving at $\sim1000\kms$ it would take them
$\sim500\Myr$ to travel to a radius of $R\sim500\kpc$ from the centre
of the cluster.  On the face of it, this is inconsistent with David \&
Kempner's proposal that the sub-clusters in the A\,1758S merger are
approximately at closest approach.  An alternative interpretation of
the $24\mic$ data might be that the shocks in A\,1758S were not strong
enough to trigger detectable star-formation in the cluster galaxies
that passed though them.  We also note that A\,1758S is much less
optically rich than A\,1758N, so the absence of obscured activity in
A\,1758S could simply be due to A\,1758S being a poorer cluster than
A\,1758N.  This is supported by Fig.~\ref{bo} because the fraction of
star-forming $24\mic$-bright galaxies in both clusters is consistent
within the errors.

In summary, our mid-IR results are qualitatively consistent with
the previous X-ray and lensing results.  However the absence of
obscured galaxies in the core of A\,1758S remains a puzzle.  One possibility is that the galaxies were already gas-poor before the cluster merger, and hence had no available fuel to undergo new star-formation. 
 Our
results confirm and extend the view formed from {\em ISO} that the
cluster-to-cluster scatter in merging clusters is large, and likely
related to the physical parameters of cluster mergers.  

\subsection{Comparison with Other Clusters}

The global SFR of A\,1758 estimated from the $24\mic$ flux of
photometrically selected members is $910{\pm}320\,\Msolpyr$,
qualifying it as one of the most infrared active clusters studied to
date.  For example it is comparable with the most active cluster
studied to date -- Cl\,0024 at $z{=}0.395$ with ${\rm
SFR}{=}1000{\pm}210\Msolpyr$ \citep{geach}.  Globally this is
unsurprising because A\,1758 contains two ongoing cluster-cluster
mergers and one (between A\,1758N and A\,1758S) that may occur within
a few Gyr.

However the picture is more subtle when considering the specific SFR
($\Sigma{\rm SFR}$) and star-forming fraction of galaxies ($f_{\rm
SF,MIPS}$ -- see Figs.~\ref{geach}~\&~\ref{bo} respectively).
Cl\,0024 stands out in Fig.~\ref{geach} as the most active cluster by
a factor of $\sim6$ in $\Sigma{\rm SFR}$ because of it's low mass
relative to A\,1758.  On the other hand, Cl\,0024 and A\,1758 have
comparable values of $f_{\rm SF,MIPS}$ (Fig.~\ref{bo}), indicative of
the clusters having comparable optical richness.  This further
underlines the conclusion that numerous parameters influence the level
of obscured star-formation in galaxy clusters: merger geometry,
density of the ICM, location of shock-heated gas with respect to the
cluster galaxies, the available supply of gas in cluster galaxies in
which star-formation may be triggered by shock-heating of the ICM, the
optical richness of the cluster, and cluster mass.  We will
investigate these issues in detail with the full sample in a future
paper.  

\subsection{Passive Spirals}

We find a significant population of passive spiral galaxies, defined
empirically by $K{<}16.5$, $n{<}2.5$ and $S_{24\mic}{<}400\uJy$.  The
radial distribution of these galaxies is intermediate in radial extent
between the cluster early-types and the actively star-forming galaxies
detected at $24\mic$ (Fig.~\ref{fig:radial}).  If the radial extent of
each distribution indicates the length of time that each population
has on average spent in the cluster potential well, then the passive
spirals have been accreted by A\,1758 less recently than the actively
star-forming galaxies.  A simple interpretation is therefore that the
passive spirals were formerly actively forming stars and thus brighter
at $24\mic$, i.e.\ that there is an evolutionary path linking the
dusty star-forming galaxies, passive spirals and cluster early-type
galaxies \citep[see e.g.][]{wolf}.

\citet{moran07} identified a significant population of passive spirals
in Cl\,0024 over a wide range of environments, and used a combination
of {\em GALEX} UV photometry and optical spectroscopy to find that
their star-formation quenching time-scales are often $>1\Gyr$,
suggestive more of starvation, although the rapid quenching expected
for ram-pressure stripping was also observed.  As the simulations of
\citet{tonnesen} show, gas is lost from infalling galaxies over a wide
range of environments (even beyond the virial radius) on both short
and long time-scales, and \citet{moran07} estimate that the observed
frequency and quenching time-scales of passive spirals in $z\sim0.4$
clusters could account for the entire build-up of S0s between $z=0.4$
and $z=0$.

The relationship between Moran et al.'s UV/optical selected passive
spirals and our optical/IR selected passive spirals is currently
unclear.  For example, the evolutionary path outlined above may not
include the UV/optical passive spirals discussed by \citet{moran07}.
On the other hand, if the two populations of passive spirals are
indeed the same class of object, then the evolutionary path may not be
valid.  In that case an alternative interpretation of the differing
radial distributions of dusty and passive spirals could be that the
physical processes that trigger star formation in the former
population are more efficient at the largest radii than the processes
that act to quench star formation in the latter population.  The {\em
GALEX}, {\em Spitzer}, {\em Herschel}, ground-based near-IR photometry
and optical spectroscopy dataset that we are assembling on a large
sample of clusters will be very powerful for disentangling the
respective galaxy populations.  

\section{Conclusions}
\label{sec:conc}

We have presented the first results from our panoramic
multi-wavelength survey of 31 clusters at $z\simeq0.2$ clusters as
part of the Local Cluster Substructure Survey (LoCuSS; PI: G.\ P.\
Smith).  For this initial study we chose A\,1758, an actively merging
cluster at $z=0.279$ that has been previously studied with high
quality weak lensing and X-ray data \citep{david,okabe}.
Specifically, we have combined a wide-field ($25'\times25'$) {\em
Spitzer}/MIPS $24\mic$ map of the entire A\,1758N/A\,1758S complex
with panoramic $J/K$-band imaging from UKIRT/WFCAM and archival
optical data from Subaru to map out the locations of dusty
star-forming galaxies in addition to cluster early-types and cluster
spirals that are undetected down to $S_{24\mic}{=}400\uJy$.  The main
results are as follows:

\smallskip

\noindent(i) We detect 82 probable cluster members in the mid-IR with
  {\em Spitzer}/MIPS at $24\mic$ within 3\,Mpc of the cluster centre.
  After correcting for residual field contamination using archival
  data from the XMM-LSS, we measure an excess of 42 mid-IR sources
  over the field counts with $S_{24\mic}{>}400\uJy$, responsible for a
  total flux of $S_{24\mic}{=}42.7{\pm}10.3\mJy$.

\smallskip

\noindent(ii) Adopting \citeauthor{lefloch}'s (\citeyear{lefloch}) SED
  templates, we convert our 90\% completeness limit of
  $S_{24\mic}{=}400\uJy$ into a luminosity sensitivity of
  $L_{8-1000\mic}{\sim}5{\times}10^{10}\Lsol$ -- i.e.\ sub-LIRG
  luminosities. Most of the IR-luminous cluster sources appear to be
  sub-LIRGs. We estimate there to be 10 LIRGs within 3\,Mpc of the
  cluster centre, comparable to the number in Cl\,0024. We identify
  just one possible cluster ULIRG ($L_{IR}{>}10^{12}\Lsol$), which
  appears to be a merging galaxy system with a combined NIR luminosity
  of ${\sim}0.4\,L_{K}^{*}$.

\smallskip

\noindent(iii) We also convert the total cluster flux at $24\mic$ to a
  total cluster star formation rate of ${\rm
  SFR_{IR}}{=}910{\pm}320\,\Msolpyr$.  This is consistent within the
  errors with the most active cluster studied to date: Cl\,0024
  \citep{geach}.  We split the $24\mic$ sources between the A\,1758N
  and A\,1758S based the proximity of each galaxy to the respective
  cluster centres, to obtain ${\rm SFR_{IR}}{=}580{\pm}210\,\Msolpyr$
  and ${\rm SFR_{IR}}{=}330{\pm}60\,\Msolpyr$ respectively -- i.e.\
  A\,1758N is almost twice as active as A\,1758S.

\smallskip

\noindent(iv) When the level of obscured activity is normalized to the
  mass and the optical richness of the cluster, then we find that the
  specific star-formation rate of A\,1758 is lower than that of
  Cl\,0024 and that the fraction of cluster galaxies that are detected
  at $24\mic$ is comparable with Cl\,0024.  These results reflect the
  facts that A\,1758 is more massive than and of comparable richness
  to Cl\,0024.

\smallskip

\noindent(v) Dust-obscured galaxies with late-type morphologies are
  detected in a network of infalling groups and filamentary structures
  that surround both A\,1758N and A\,1758S that is delineated by the
  reconstructed cluster mass distribution from \citet{okabe}.  However
  in the core regions of each cluster we find that the central
  $R{\ls}500\,\kpc$ region of A\,1758N is comparatively rich in dusty
  galaxies in contrast to the absence of such objects in the same
  region of A\,1758S.  This difference may be due to the differing
  recent dynamical histories of the two clusters, as discussed by
  \cite{david}.

\smallskip

\noindent(vii) Finally, we identify a population of IR-selected
  passive spiral galaxies, i.e.\ they have a late-type morphology and
  are undetected at $24\mic$.  The radial distribution of these
  galaxies does not extend as far as that of the MIPS-detected
  population, suggesting that they were accreted into the cluster
  potential on average longer ago than the active dusty population.
  However the difference in radial distribution may also indicate that
  they are a distinct population in transition from spiral to
  early-type morphology, and that they are affected by different
  physical process than the dusty galaxies.  Comparison of these IR
  passive spirals with UV-selected passive spirals will help to
  clarify the status of both populations.

\smallskip

The next steps in this study are to repeat this analysis for a large
(${\sim}2$0--30) sample of clusters in order to measure for the first
time the scatter in the infrared properties of clusters.  The biggest
caveat on our results is the reliance on photometry to select cluster
galaxies.  We will therefore also significantly enhance the precision
of our analysis as spectroscopic redshifts become available from our
ground-based optical spectroscopic observations recently commenced
with Hectospec at the MMT.  We also plan to integrate this
optical/near-IR/mid-IR analysis with our {\em GALEX} observations of
the same clusters, and {\em Herschel} data when they become available.
The overall goal is to pinpoint the contribution that the various
physical processes make to the transformation of gas rich spiral
galaxies into early-type cluster galaxies via a large and detailed
study of clusters in a single redshift slice.

\section*{acknowledgements}

CPH and GPS acknowledge financial support from STFC.  GPS and RSE acknowledge
support from the Royal Society.  
We acknowledge NASA funding for this project under the Spitzer
program GO:40872. We wish to thank the staff at UKIRT, CASU and WFAU for making the observations and rapidly processing the NIR data.

\label{lastpage}

\begin{thebibliography}{99}
\bibitem[\protect\citeauthoryear{Bai et al.}{2007}]{bai}
 Bai L., et al., 2007, ApJ, 664, 181
\bibitem[\protect\citeauthoryear{Baldwin et al.}{1981}]{bpt}
 Baldwin A., Phillips M. M., Terlevich R., 1981, PASP, 93, 5
\bibitem[\protect\citeauthoryear{Ball et al.}{2008}]{ball}
 Ball N. M., Loveday J., Brunner R. J., 2008, MNRAS, 383, 907
\bibitem[\protect\citeauthoryear{Barger et al.}{1996}]{barger}
 Barger A. J., Aragon-Salamanca A., Ellis R. S., Couch W. J., Smail I., Sharples R. M., 1996, MNRAS, 279, 1
\bibitem[\protect\citeauthoryear{Bertin \& Arnouts}{1996}]{bertin}
 Bertin E., \& Arnouts S., 1996, A\&AS, 117, 393
\bibitem[\protect\citeauthoryear{Biviano et al.}{2004}]{biviano}
 Biviano A., et al., 2004, A\&A, 425, 33
\bibitem[\protect\citeauthoryear{B\"{o}hringer et al.}{2004}]{bohringer}
 B\"{o}hringer H., et al., 2004, A\&A, 425, 367
\bibitem[\protect\citeauthoryear{Boselli \& Gavazzi}{2006}]{boselli}
 Boselli A., \& Gavazzi G., 2006, PASP, 118, 517
\bibitem[\protect\citeauthoryear{Bruzual \& Charlot}{2003}]{bc03}
 Bruzual G., \& Charlot S., 2003, MNRAS, 344, 1000
\bibitem[\protect\citeauthoryear{Butcher \& Oemler}{1978}]{bo78}
 Butcher H., \& Oemler Jr. A., 1978, ApJ, 219, 18 
\bibitem[\protect\citeauthoryear{Butcher \& Oemler}{1984}]{bo84}
 Butcher H., \& Oemler Jr. A., 1984, ApJ, 284, 426 
\bibitem[\protect\citeauthoryear{Caldwell et al.}{1993}]{caldwell}
 Caldwell N., Rose J. A., Sharples R. M., Ellis R. S., Bower R. G., 1993, AJ, 106 473
\bibitem[\protect\citeauthoryear{Calzetti et al.}{2007}]{calzetti}
 Calzetti D., et al., 2007, ApJ, 666, 870
\bibitem[\protect\citeauthoryear{Casali et al.}{2007}]{wfcam}
 Casali M., et al., 2007, A\&A, 467, 777
\bibitem[\protect\citeauthoryear{Coia et al.}{2005a}]{coia05a}
 Coia D., et al., 2005a, A\&A, 430, 59
\bibitem[\protect\citeauthoryear{Coia et al.}{2005b}]{coia05b}
 Coia D., et al., 2005b, A\&A, 431, 433
\bibitem[\protect\citeauthoryear{Cowie et al.}{2004}]{cowie}
 Cowie L. L., Barger A. J., Fomalont E. B., Capak P., 2004, ApJL, 603, 69
\bibitem[\protect\citeauthoryear{Crowl \& Kenney}{2008}]{crowl}
 Crowl H. H., \& Kenney J. D. P., 2008, AJ, 136, 1623
\bibitem[\protect\citeauthoryear{Czoske et al.}{2002}]{czoske}
 Czoske O., Moore B., Kneib J.-P., Soucail G., 2002, A\&A, 386, 31
\bibitem[\protect\citeauthoryear{David \& Kempner}{2004}]{david}
 David L. P., Kempner J., 2004, ApJ, 613, 831
\bibitem[\protect\citeauthoryear{De Propris et al.}{2007}]{depropris}
De Propris R., Stanford S. A., Eisenhardt P. R., Holden B. P., Rosati P., 2007, AJ, 133, 2209
\bibitem[\protect\citeauthoryear{Desai et al.}{2007}]{desai}
 Desai V., et al., 2007, ApJ, 660, 1151
\bibitem[\protect\citeauthoryear{Dressler}{1980}]{dressler80}
 Dressler A., 1980, ApJ, 236, 351
\bibitem[\protect\citeauthoryear{Dressler et al.}{1997}]{dressler97}
 Dressler A., et al., 1997, ApJ, 490, 577
\bibitem[\protect\citeauthoryear{Dressler et al.}{2008}]{dressler08}
 Dressler A., Rigby J., Oemler A. Jr., Fritz J., Poggianti B., Rieke G., Bai L., 2008, arXiv/0806.2343
\bibitem[\protect\citeauthoryear{Driver et al.}{2006}]{driver}
 Driver S., et al., 2006, MNRAS, 368, 41
\bibitem[\protect\citeauthoryear{Duc et al.}{2004}]{duc04}
 Duc P.-A., et al., 2004, in IAU Colloq. 195: Outskirts of Galaxy Clusters: Intense Life in the Suburbs, ed. A. Diaferio, 347--351
\bibitem[\protect\citeauthoryear{Duc et al.}{2002}]{duc02}
 Duc P.-A., et al., 2002, A\&A, 382, 60
\bibitem[\protect\citeauthoryear{Ebeling et al.}{1998}]{ebeling98}
 Ebeling H., Edge A. C., Bohringer H., Allen S. W., Crawford C. S., Fabian A. C., Voges W., Huchra J. P., 1998, MNRAS, 301, 881
\bibitem[\protect\citeauthoryear{Ebeling et al.}{2000}]{ebeling00}
 Ebeling H., Edge A. C., Allen S. W., Crawford C. S., Fabian A. C., Huchra J. P., 2000, MNRAS, 318, 333
\bibitem[\protect\citeauthoryear{Egami et al.}{2006}]{egami}
 Egami E., et al., 2006, ApJ, 647, 922
\bibitem[\protect\citeauthoryear{Ellingson et al.}{2001}]{ellingson}
 Ellingson E., Lin H., Yee H. K. C., Carlberg R. G., 2001, ApJ, 547, 609
\bibitem[\protect\citeauthoryear{Fadda et al.}{2000}]{fadda}
 Fadda D., Elbaz D., Duc P.-A., Flores H., Franceschini A., Cesarsky C. J., Moorwood A. F. M., 2000, A\&A, 361, 827
\bibitem[\protect\citeauthoryear{Gallazzi et al.}{2009}]{gallazzi}
 Gallazzi A., et al., 2009, ApJ, 690, 1883
\bibitem[\protect\citeauthoryear{Geach et al.}{2006}]{geach}
 Geach J.E., Smail I., Ellis R. S., Moran S. M., Smith G. P., Treu T., Kneib J.-P., Edge A. C., Kodama A. C., 2006, ApJ, 649, 661 
\bibitem[\protect\citeauthoryear{Gehrels}{1986}]{gehrels}
 Gehrels N., 1986, ApJ, 303, 336
\bibitem[\protect\citeauthoryear{Gordon et al.}{2005}]{gordon}
 Gordon K. D., et al., 2005, PASP, 117, 505
\bibitem[\protect\citeauthoryear{Haines et al.}{2007}]{haines07}
 Haines C. P., Gargiulo A., La Barbera F., Mercurio A., Merluzzi P., Busarello G., 2007, MNRAS, 381, 7
\bibitem[\protect\citeauthoryear{Haines et al.}{2009}]{haines09}
 Haines C. P., et al., 2009, in preparation
\bibitem[\protect\citeauthoryear{Hickox et al.}{2009}]{hickox}
 Hickox R. C., et al., 2009, preprint (arXiv:0901.4121)
\bibitem[\protect\citeauthoryear{Holden et al.}{2007}]{holden}
 Holden B. P., et al. 2007, ApJ, 670, 190
\bibitem[\protect\citeauthoryear{Kauffmann}{1995}]{kauffmann95}
 Kauffmann G., 1995, MNRAS, 274, 153
\bibitem[\protect\citeauthoryear{Kennicutt et al.}{2007}]{kennicutt07}
 Kennicutt R. C. Jr., et al. 2007, ApJ, 671, 333
\bibitem[\protect\citeauthoryear{Kneib et al.}{1996}]{kneib}
 Kneib J.-P., Ellis R. S., Smail I., Couch W. J., Sharples R. M., 1996, ApJ, 471, 643
\bibitem[\protect\citeauthoryear{La Barbera et al.}{2008}]{labarbera}
 La Barbera F., et al. 2008, PASP, 120, 681
\bibitem[\protect\citeauthoryear{Lawrence et al.}{2007}]{ukidss}
 Lawrence A. et al. 2007, MNRAS, 379, 1599
\bibitem[\protect\citeauthoryear{Le F\`{e}vre et al.}{2005}]{vvds}
 Le F\`{e}vre O. et al. 2005, A\&A, 439, 640
\bibitem[\protect\citeauthoryear{Le Floc'h et al.}{2005}]{lefloch}
 Le Floc'h E., et al. 2005, ApJ, 632, 169
\bibitem[\protect\citeauthoryear{Lewis et al.}{2002}]{lewis}
 Lewis I. et al. 2002, MNRAS, 334, 673  
\bibitem[\protect\citeauthoryear{Limousin et al.}{2007}]{limousin}
 Limousin M., et al., 2007, ApJ, 668, 643
\bibitem[\protect\citeauthoryear{Lonsdale et al.}{2003}]{swire}
 Lonsdale C. et al. 2003, PASP, 115, 897
\bibitem[\protect\citeauthoryear{Marcillac et al.}{2007}]{marcillac}
 Marcillac D., Rigby J. R., Rieke G. H., Kelly D. M., 2007, ApJ, 654, 825
\bibitem[\protect\citeauthoryear{Martig \& Bournaud}{2005}]{martig}
 Martig M., \& Bournaud F., 2008, MNRAS, L38
\bibitem[\protect\citeauthoryear{Martini et al.}{2006}]{martini}
 Martini P., Kelson D. D., Kim E., Mulchaey J. S., Athey A. A., 2006, ApJ, 644, 116
\bibitem[\protect\citeauthoryear{Metcalfe et al.}{2003}]{metcalfe03}
 Metcalfe L., et al., 2003, A\&A, 407, 791
\bibitem[\protect\citeauthoryear{Metcalfe et al.}{2005}]{metcalfe}
 Metcalfe L., Fadda D., \& Biviano A., 2005, SSRv, 119, 425
\bibitem[\protect\citeauthoryear{Miller et al.}{2003}]{miller03}
 Miller N. A., \& Owen F. N., 2003, AJ, 125, 2427
\bibitem[\protect\citeauthoryear{Miller et al.}{2006}]{miller06}
 Miller N. A., Oegerle W. R., \& Hill J. M., 2006, AJ, 131, 2426
\bibitem[\protect\citeauthoryear{Miyazaki et al.}{2002}]{suprime}
 Miyazaki S., et al., 2002, PASJ, 54, 833
\bibitem[\protect\citeauthoryear{Moran et al.}{2005}]{moran05}
 Moran S. M., Ellis R. S., Treu T.,  Smail I., Dressler A., Coil A., Smith G. P., 2005, ApJ, 634, 977
\bibitem[\protect\citeauthoryear{Moran et al.}{2007}]{moran07}
 Moran S. M., Ellis R. S., Treu T., Smith G. P., Rich M. R., Smail I., 2007, ApJ, 671, 1503
\bibitem[\protect\citeauthoryear{Moss}{2006}]{moss}
 Moss C., 2006, MNRAS, 373, 167
\bibitem[\protect\citeauthoryear{Muzzin et al.}{2008}]{muzzin}
 Muzzin A., Wilson G., Lacy M., Yee H. K. C., Stanford S. A., 2008, ApJ, 686, 966
\bibitem[\protect\citeauthoryear{Navarro, Frenk \& White}{1997}]{nfw}
 Navarro J. F., Frenk C. S., \& White S. D. M., 1997, ApJ, 490,493
\bibitem[\protect\citeauthoryear{Okabe \& Umetsu}{2008}]{okabe}
 Okabe N., \& Umetsu K., 2008, PASJ, 60, 3450
\bibitem[\protect\citeauthoryear{Okabe et al.}{2009}]{okabe09}
 Okabe N., Takada M., Umetsu K., Futamase T., Smith G. P., 2009, preprint (arXiv:0903.1103)
\bibitem[\protect\citeauthoryear{Owen et al.}{1999}]{owen}
 Owen F., Ledlow M. J., Keel W. C., Morrison G. E., 1999, AJ, 118, 633
\bibitem[\protect\citeauthoryear{Poggianti et al.}{1999}]{poggianti}
 Poggianti B. M., Smail I., Dressler A., Couch W. J., Barger A. J., Butcher H., Ellis R. S., Oemler A. Jr., 1999, ApJ, 518, 576
\bibitem[\protect\citeauthoryear{Poggianti et al.}{2001}]{poggianti01}
 Poggianti B. M., Bressan A., Franceschini A., 2001, ApJ, 550, 195
\bibitem[\protect\citeauthoryear{Postman et al.}{2005}]{postman}
 Postman M., et al., 2005, ApJ, 623, 721
\bibitem[\protect\citeauthoryear{Rieke et al.}{2004}]{rieke}
 Rieke G. H., et al., 2004, ApJS, 154, 25
\bibitem[\protect\citeauthoryear{Rizza et al.}{2003}]{rizza}
 Rizza E., Morrison G. E., Owen F. N., Ledlow M. J., Burns J. O., Hill J., 2003, AJ, 126, 119
\bibitem[\protect\citeauthoryear{Roettiger, Burns \& Loken}{1996}]{roettiger}
 Roettiger K., Burns J. O., \& Loken C., 1996, ApJ, 473, 651
\bibitem[\protect\citeauthoryear{Saintonge et al.}{2008}]{saintonge}
 Saintonge A., Tran K.-V. H., Holden B. P., 2008, ApJL, 685, 113
\bibitem[\protect\citeauthoryear{Silverman et al.}{2008}]{silverman}
 Silverman J. D., et al., 2008, preprint (arXiv:0812.3402)
\bibitem[\protect\citeauthoryear{Sivakoff et al.}{2008}]{sivakoff}
 Sivakoff G., Martini P., Zabludoff A. I., Kelson D. D., Mulchaey J. S., ApJ, 682, 803
\bibitem[\protect\citeauthoryear{Smail et al.}{1999}]{smail}
 Smail I., et al., 1999, ApJ, 525, 609 
\bibitem[\protect\citeauthoryear{Smith et al.}{2005a}]{smith}
 Smith G. P., Treu T., Ellis R. S., Moran S. M., Dressler A., 2005, ApJ, 620, 78
\bibitem[\protect\citeauthoryear{Smith et al.}{2005b}]{smith05}
 Smith G. P., Kneib J.-P., Smail I., Mazzotta P., Ebeling H., Czoske O., 2005, MNRAS, 359, 417
\bibitem[\protect\citeauthoryear{Smith et al.}{2008}]{smith08}
 Smith G. P., \& Taylor J. E., 2008, ApJL, 682, 73
\bibitem[\protect\citeauthoryear{Swinbank et al.}{2006}]{swinbank}
 Swinbank A. M., Bower R. G., Smith G. P., Smail I., Kneib J.-P., Ellis R. S., Stark D. P., Bunker A. J., 2006, MNRAS, 368, 1631
\bibitem[\protect\citeauthoryear{Tanaka et al.}{2005}]{tanaka}
Tanaka M., Kodama T., Arimoto N., Okamura S., Umetsu K., Shimasaku K., Tanaka I., Yamada T., 2005, MNRAS, 362, 268
\bibitem[\protect\citeauthoryear{Tomita et al.}{1996}]{tomita}
 Tomita A., Nakamura F. E., Takata T., Nakanishi K., Takeuchi T., Ohta K., 1996, AJ, 111, 42
\bibitem[\protect\citeauthoryear{Tonnesen, Bryan \& van Gorkum}{2007}]{tonnesen}
 Tonnesen S., Bryan G. L., \& van Gorkum J. H., 2007, ApJ, 671, 1434
\bibitem[\protect\citeauthoryear{Tremonti et al.}{2004}]{tremonti}
 Tremonti C. et al., 2004, ApJ, 613, 898
\bibitem[\protect\citeauthoryear{Treu et al.}{2003}]{treu}
 Treu T., et al. 2003, ApJ, 591, 53
\bibitem[\protect\citeauthoryear{Werner et al.}{2004}]{werner}
 Werner M. W., et al., 2004, ApJS, 154, 1
\bibitem[\protect\citeauthoryear{Wolf et al.}{2009}]{wolf}
 Wolf C., et al., 2009, MNRAS, 393, 1302
\bibitem[\protect\citeauthoryear{Zhang}{2008}]{zhang}
 Zhang X., 2008, PASP, 120, 121
\bibitem[\protect\citeauthoryear{Zhang et al.}{2008}]{zhang08}
 Zhang Y.-Y., Finoguenov A., B\"{o}hringer H., Kneib J.-P., Smith G. P., Kneissl R., Okabe N., Dahle H., 2008, A\&A, 482, 451
\end{thebibliography}
\end{document}